\documentclass[structabstract]{aa}  
\usepackage{graphicx}
\usepackage{natbib}
\usepackage{txfonts}
\begin{document}
\def\frac{$''$\hspace*{-.1cm}}
\def\deg{$^{\circ}$}
\def\min{$'$}
\def\deg{$^{\circ}$\hspace*{-.1cm}}
\def\hii{H\,{\sc ii}}
\def\hi{H\,{\sc i}}
\def\hg{H$\gamma$}
\def\hd{H$\delta$}
\def\he{H$\epsilon$}
\def\hb{H$\beta$}
\def\ha{H$\alpha$}

\def\lam{$\lambda$}

\def\aiii{Al~{\sc{iii}}\ }
\def\ariii{[Ar\,{\sc{iii}}]}
\def\ariv{[Ar\,{\sc{iv}}]}
\def\cii{[C\,{\sc{ii}}]}
\def\ciii{C\,{\sc{iii}}}
\def\civ{C\,{\sc{iv}}}
\def\caii{Ca\,{\sc{ii}}}
\def\cliii{[Cl\,{\sc{iii}}]}
\def\crii{Cr\,{\sc{ii}}}
\def\fei{Fe\,{\sc{i}}}
\def\feii{Fe\,{\sc{ii}}}
\def\feiii{Fe\,{\sc{iii}}}
\def\hei{He\,{\sc{i}}}
\def\heii{He\,{{\sc ii}}}
\def\nii{[N\,{\sc{ii}}]}
\def\niii{N\,{\sc{iii}}}
\def\niv{N\,{\sc{iv}}}
\def\nv{N\,{\sc{v}}}
\def\ni{Na~{\sc{i}}\ }
\def\nei{Ne~{\sc{i}}\ }
\def\neiii{[Ne~{\sc{iii}}]}
\def\oi{[O\,{\sc i}]}
\def\oii{[O\,{\sc ii}]}
\def\oiii{[O\,{\sc iii}]}
\def\sii{[S\,{\sc ii}]}
\def\siii{[S\,{\sc iii}]}
\def\siv{S\,{\sc{iv}}}
\def\si_ii{Si\,{\sc{ii}}}
\def\si_iii{Si\,{\sc{iii}}}
\def\siiv{Si\,{\sc{iv}}}
\def\srii{Sr\,{\sc ii}}
\def\tiii{Ti\,{\sc{ii}}}
\def\yii{Y\,{\sc ii}}
\def\kv{K\,{\sc v}}

\def\x{$\times$}
\def\av{$A_{V}$}
\def\mcube{$^{-3}$}
\def\cm2{cm$^{-2}$}
\def\sec{s$^{-1}$}

\def\sm{$M_{\odot}$}
\def\slum{$L_{\odot}$}
\def\ab{$\sim$}
\def\sec{s$^{-1}$}

\title{Two compact \hii\ regions at the remote outskirts of the Magellanic Clouds\thanks{Based 
on observations obtained at the European Southern 
   Observatory, La Silla, Chile, Program 69.C-0286(A) and 69.C-0286(B).}}

\author{R. Selier\inst{1} \and M. Heydari-Malayeri\inst{1}
}

\institute{Laboratoire d'Etudes du Rayonnement et de la Mati\`ere en Astrophysique (LERMA), 
Observatoire de Paris, CNRS, \\ 
61 Avenue de l'Observatoire, 75014 Paris, France,  Romain.Selier@obspm.fr
}

\authorrunning{Selier et al.}
\titlerunning{}

\date{Received * / accepted *}

  \abstract
   {}
  {The \hii\ regions LMC N191 and SMC N77 are among the outermost massive 
star-forming regions in the Magellanic Clouds. So far, few works have dealt with these objects 
despite their interesting characteristics. We aim at studying various physical properties 
of these objects regarding their morphology (in the optical and Spitzer IRAC wavelengths), 
 ionized gas emission, 
nebular chemical abundances, exciting sources, stellar content, age, presence or absence of  
 young stellar objects, etc. 
}
   {This study is based mainly on optical ESO NTT observations,  
both imaging and spectroscopy, coupled with other archive data, notably Spitzer images 
(IRAC 3.6, 4.5, 5.8, and 8.0\,$\mu$m) and  2MASS observations.
} 
   { We show the presence of two compact \hii\ regions, a low-excitation blob (LEB) named LMC N191A 
and a high-excitation
blob (HEB) named SMC N77A, and study their properties 
and those of their exciting  
massive stars as far as  spectral type and mass are concerned.  
We also analyze the environmental stellar populations and determine their evolutionary 
stages. Based on Spitzer IRAC data, we characterize the YSO candidates detected in the 
direction of these regions. Massive star formation is going 
on in these young regions with protostars of mass \ab\,10 and 20 \sm\  
in the process of formation.
}
   {}


   \maketitle
%

\section{Introduction}

The compact \hii\ regions residing in the Magellanic Clouds 
are interesting in the context of massive star formation in these neighboring 
galaxies.  Typical Magellanic Cloud \hii\ regions are giant complexes of ionized gas 
with sizes of several arc minutes, corresponding to physical scales of more than 
50\,pc and are powered by a large number of exciting stars. In contrast, 
Magellanic Cloud compact \hii\ regions are small 
regions mostly $\sim$\,5\frac\, to 10\frac\, in diameter, 
corresponding to $\sim$\,1.5 to 3.0\,pc and excited by a much smaller number 
of massive stars. There are two types of compact \hii\ regions, 
high-excitation blobs \citep[HEBs, for a review see][]{MHM10b} 
and low-excitation blobs \citep[LEBs,][]{Meynadier07}. 
The members of the first group are often observed lying adjacent or projected onto  
giant \hii\ regions and are younger than the associated giant \hii\ regions. 
Do HEBs indeed belong 
to the same region of the Magellanic Clouds at which the giant \hii\ regions have formed or 
is the association between these two types of \hii\ regions a line-of-sight effect? 
If they are associated, are HEBs powered by triggered, second-generation massive 
stars?  Why has star formation not proceeded in a single burst although 
massive stars are believed to form in the dense core of giant molecular clouds? 
These are some interesting questions, the answers to which  
will be helpful for better understanding massive star formation in the Magellanic Clouds. 
A problem is that HEBs are not numerous, and moreover, few of them have been 
studied individually in detail. \\

This paper is devoted to a first detailed study of two compact \hii\ regions, one in 
the Large Magellanic Cloud (LMC) \hii\ region N191 and the 
other in the Small Magellanic Cloud (SMC) N77 
\citep[][]{Henize56}.  
Among the LMC \hii\ regions listed by \citet[][]{Henize56}, N191 is one of the 
outermost, lying below the bar, at a distance of \ab\ 200\min\ (\ab\ 3 kpc in projection) 
from the famous 30 Doradus. N191 appears as an elongated structure, 
with two components N191A and N191B in the Henize catalog. 
Here we essentially study the brightest component N191A, also known as 
DEM L 64b \citep[][]{Davies76}.
N77 is one of the most northern \hii\ regions of the 
SMC; it is situated at a distance of \ab\ 25\min\ (\ab\ 440 pc in projection) 
from the pre-eminent SMC 
\hii\ region N66 \citep[][and references therein]{MHM10a}. 
SMC N77 is identified in the optical survey of 
\citet[][]{Davies76} as DEM S 117. \\

Few works have been devoted to these two \hii\ regions despite their interesting characteristics. 
LMC N191 belongs to the OB association LH 23 \citep[][]{Lucke70}. It was also detected as 
IRAS source 05051-7058 \citep[][]{Helou88}.
The compact \hii\ region SMC N77 seems to coincide with the stellar association B-OB 24 
\citep[]{Battinelli91}. It was identified in the infrared  as IRAS source 01011-7209 
\citep[][]{Helou88} and as source \#48 in the ISO 12\,$\mu$m catalog \citep[][]{Wilke03}.
Furthermore, LMC N191 and SMC N77 were part of a Spitzer study of compact \hii\ regions 
by \citet[][]{Charmandaris08} and have been included in several radio continuum surveys of 
the Magellanic Clouds \citep[][]{Filipovic95,Filipovic02}.
Both compact \hii\ regions are associated with molecular clouds. 
The giant molecular cloud LMC N J0504-7056 is centered at 130\frac\ south of N191 
\citep[][]{Fukui08}. Moreover, the OB association LH 23 and the \hii\ region are related to 
this molecular cloud \citep[][]{Fukui08,Kawamura09}. 
A small molecular cloud has been detected near the position of SMC N77 
\citep[][]{Mizuno01}.  \\

This paper is arranged as follows. Section 2 presents the observations, 
data reduction, and the archive data (Spitzer data, 2MASS data). 
Section 3 describes our results (overall view, extinction, nebular emission, 
stellar content and chemical abundances). 
Section 4 presents our discussion, and finally our conclusions are summarized in Section 5.

\section{Observations and data reduction}

\subsection{NTT imaging}

LMC N191 and SMC N77 were observed on 28 September 2002 using the
ESO New Technology Telescope (NTT) equipped with the active optics and
the Superb Seeing Imager \citep[SuSI2;][]{D'Odorico98}. 
The detector consisted of two CCD
chips, identified as ESO \#45 and \#46.  The two resulting frames were 
automatically combined to produce a single FITS file, while the space
between the two chips was ``filled'' with some overscan columns so that
the respective geometry of the two chips was approximately
preserved. The gap between the chips corresponds to \ab\,100 true CCD
pixels, or \ab\,8\frac.  The file format was 4288\,\x\,4096
pixels, and the measured pixel size  0\frac.0805 on the sky. Each
chip of the mosaic covered a field of 5\min.5\,\x\, 2\min.7. We refer to the
ESO manual SuSI2 for more technical information.  \\

Nebular imaging was carried out using the narrow-band filters centered
on the emission lines \ha\, (ESO \#884), \hb\, (\#881), and
\oiii\,(\#882). N191 was observed with three exposures of 180 sec for each filter. 
N77 was also observed with exposures of 180 sec: five exposures in \ha\ and three 
\hb\ and \oiii\ exposures. The image quality was quite good during the night,  
the seeing was 0\frac.8.   
We constructed the line-ratio maps \ha/\hb\, and  \oiii/\hb\  
from nebular imaging. We also took exposures using filters ESO
\#811 ($B$), \#812 ($V$), and \#813 ($R$) with unit 
exposure times of 15 sec for $B$ and $V$ and 10 sec for $R$,  
respectively. The exposures for each filter were repeated twice  
using ditherings of 5\frac\,--10\frac\, for bad pixel 
rejection. \\

PSF-fitting photometry was obtained for all filters using the DAOPHOT package under 
IRAF\footnote{http://iraf.noao.fr}. The magnitudes were then calibrated using the photometric 
calibration package photcal. To perform this calibration, seven standard stars, 
belonging to two Landolt photometric groups SA\,92 and T\,Phe 
\citep[][]{Landolt92} were observed at 
four different airmasses. This led to the determination of the photometry coefficients 
and zero-points. Those coefficients agree well with the indicative 
values displayed on the SuSI2 web page.\\

The aperture corrections were calculated as 
follows. Starting from one of the frames, we subtracted  
all stars except those used for determining the PSF 
with the daophot.substar procedure, using our preliminary 
DAOPHOT photometry and the corresponding PSF. This led 
to a frame with only a few bright, isolated stars plus residues 
from the subtraction. We then performed both aperture and 
PSF-fitting photometry on those stars, using the same aperture 
as for standard stars. The comparison led to aperture corrections  
of 0.02, 0.04, and 0.03 mag in $B$, $V$, and $R$, respectively.\\

During the photometry process, some slight discrepancies 
between the intensity of the frames were found: this effect was 
considered to be the consequence of episodic variations in the sky 
transparency by 7$\%$ at most. To avoid introducing a systematic 
underestimation of star magnitudes when averaging the frames, we decided to 
perform photometry on each individual frame.\\

By cross-correlating the positions of the sources in the various photometry 
files, we obtained the mean magnitude (average of the 2 mag of each filter) and a 
decent estimator of the uncertainty in this magnitude (difference between maximum 
and minimum magnitudes). Finally, the process yielded the 
photometry of 644 stars for the LMC N191 field and 
236 stars for that of SMC N77 in all 
three filters. This difference of the number of sources is partly due to the limit magnitude 
of the photometry (\ab\ 21 mag for LMC N191, \ab\ 20 mag for SMC N77). 
It is better for LMC N191 than for SMC N77 because of the better sky conditions.
The results for the brightest stars 
toward LMC N191 and SMC N77 are presented in Table\,\ref{tab:stars_n191}. The whole photometry is 
available in electronic form.

\subsection{NTT spectroscopy}

The EMMI spectrograph \citep[][]{Dekker86}  
attached to the ESO NTT telescope was used on 29  
September 2002 to obtain several long-slit stellar spectra. 
The grating was \#\,12 centered on 4350\,\AA\, (BLMRD mode)
and the detector was a Tektronix CCD TK1034 with 1024$^{2}$ pixels of
size 24 $\mu$m.  The covered wavelength range was 3810-4740\,\AA\, 
and the dispersion 38\,\AA\,mm$^{-1}$, giving {\sc fwhm} 
resolutions of $2.70\pm0.10$ pixels or $2.48\pm0.13$\,\AA\, for a 1\frac.0 slit. 
At each position, we took three 10-min exposures. The instrument response was derived 
from observations of the calibration stars  LTT\,7379, LTT\,6248, and  LTT\,7987. 
The seeing condition was 0\frac.8 ({\sc fwhm}). The identifications of the stars 
along the slits 
were based on monitor sketches drawn during the observations.  \\

Furthermore, EMMI was used on 28 September 2002 to obtain nebular 
spectra with gratings \#\,8 (4550-6650\,\AA) and \#\,13 4200-8000) in the REMD mode 
and with grating \#\,4 (3650-5350\,\AA) in the BLMD mode. In the REMD mode,    
the detector was  CCD \#\,63, 
MIT/LL, 2048\,\x\,4096 pixels of  15$^{2}\,\mu$m$^{2}$ each.   
Spectra were obtained with the slit set in 
east-west and north-south orientations using a basic exposure time 
of 300 sec repeated several times. The seeing conditions varied around 
0\frac.7. 
Reduction and extraction of spectra were performed using the IRAF software
package. Fluxes were derived from the extracted spectra with the 
IRAF task SPLOT. The line fluxes were measured by fitting 
Gaussian profiles to the lines as well as by simple pixel integration in 
some cases. The nebular line intensities were corrected for interstellar 
reddening using the formulae given by \citet{Howarth83} for the LMC extinction,  
which is very similar to that of the SMC in the visible. 
The intensities of the main nebular lines
are presented in Table \ref{tab:flux} where $F(\lambda)$ and $I(\lambda)$ represent
observed and de-reddened line intensities. The uncertainties are
indicated by capital letters: A $<$\,10\%, B=10--20\%, C=20--30\%, and D
$>$\,30\%. \\

\subsection{Archive Spitzer and 2MASS data}

We used data obtained with the Infrared Array Camera (IRAC) 
on board the Spitzer Space Telescope to build composite images of 
LMC N191 and SMC N77 and also to carry out their photometry.
The observations of the LMC were part of the SAGE-LMC survey 
(PI M. Meixner, PID = 20 203, see \citet[][]{Meixner06}), while the SMC 
data belong to the S$^{3}$MC
project (PI A. Bolatto PID = 3316, see \citet[][]{Bolatto07}). \\

The typical PSF of the IRAC images in the 
3.6, 4.5, 5.8, and 8.0 $\mu$m bands is 1\frac.66 to  1\frac.98. 
The derived photometry for LMC N191 in the 3.6, 4.5, 5.8, and 8.0 $\mu$m 
bands are 10.50, 9.48, 8.03, and 6.47 mag \citep{Charmandaris08}, 
 respectively, using an integration 
aperture of 3 pixels, or 3.6\frac\, in radius 
\citep{Charmandaris08}. Using the same aperture the IRAC photometry for 
SMC N77 in the 3.6, 4.5, 5.8, and 8.0 $\mu$m 
bands are 13.86, 13.87, 12.65, and 10.64 mag \citep{Charmandaris08}. Measurements with 
either slightly larger or smaller apertures do not affect the color results. \\

We also used the {\it JHK} photometry provided by the 2MASS point source catalog 
(http://tdc-www.harvard.edu/catalogs/tmpsc.html), as presented in  
Table\,\ref{tab:stars_n191}. 
Note that the embedded stars in the \hii\ region LMC N191A  
(i.e. \#1, \#2, \#3, \#4 and \#5) are not resolved in 2MASS data, so 
the {\it JHK} photometry of the star N191-1 corresponds to the whole N191A 
compact \hii\ region. The same is true for the {\it JHK} photometry 
of SMC N77-1, which corresponds to the whole N77A \hii\ region.

\begin{figure*}[]
\centering

\includegraphics[width=0.7\hsize]{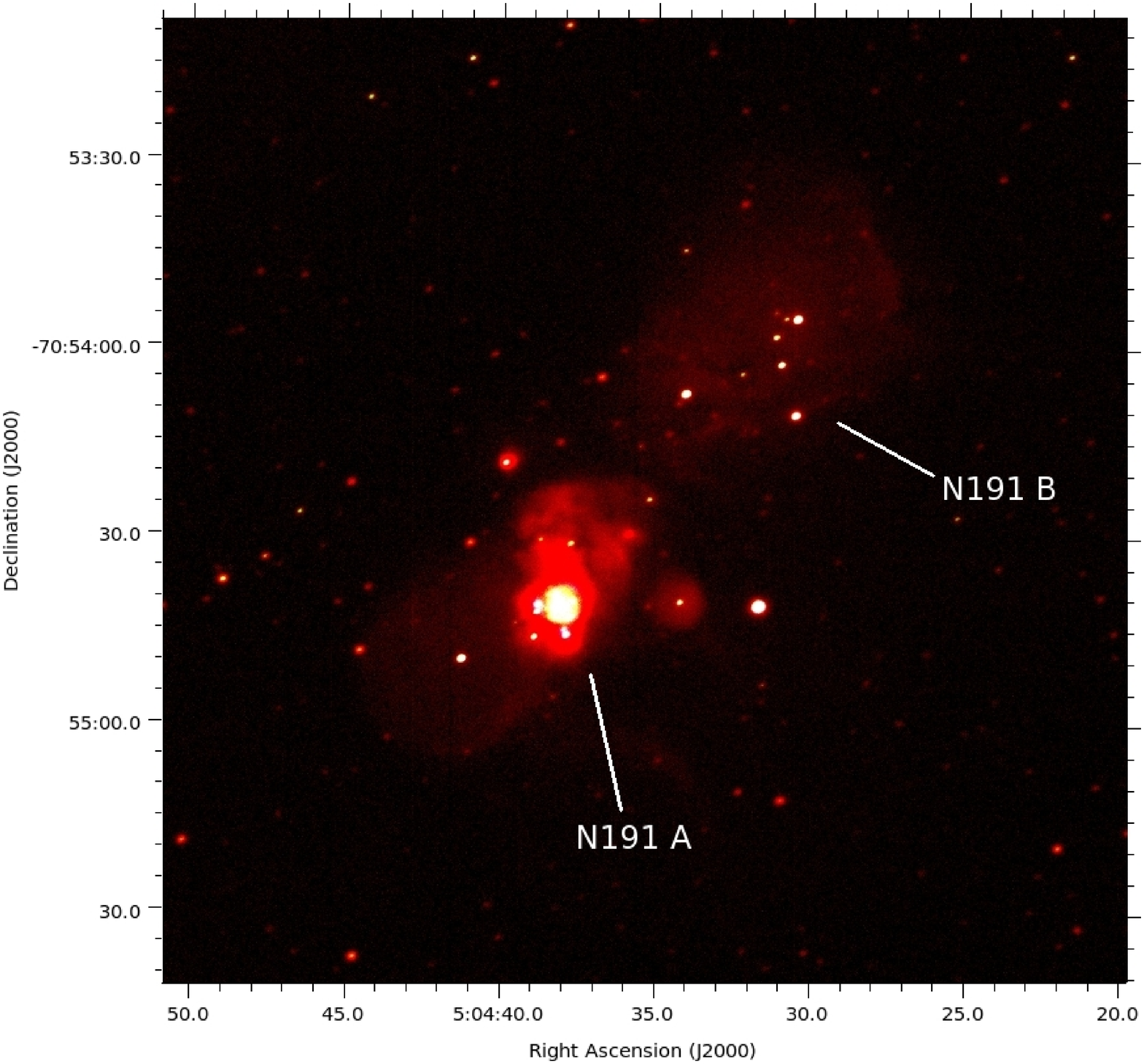}
\includegraphics[width=0.7\hsize]{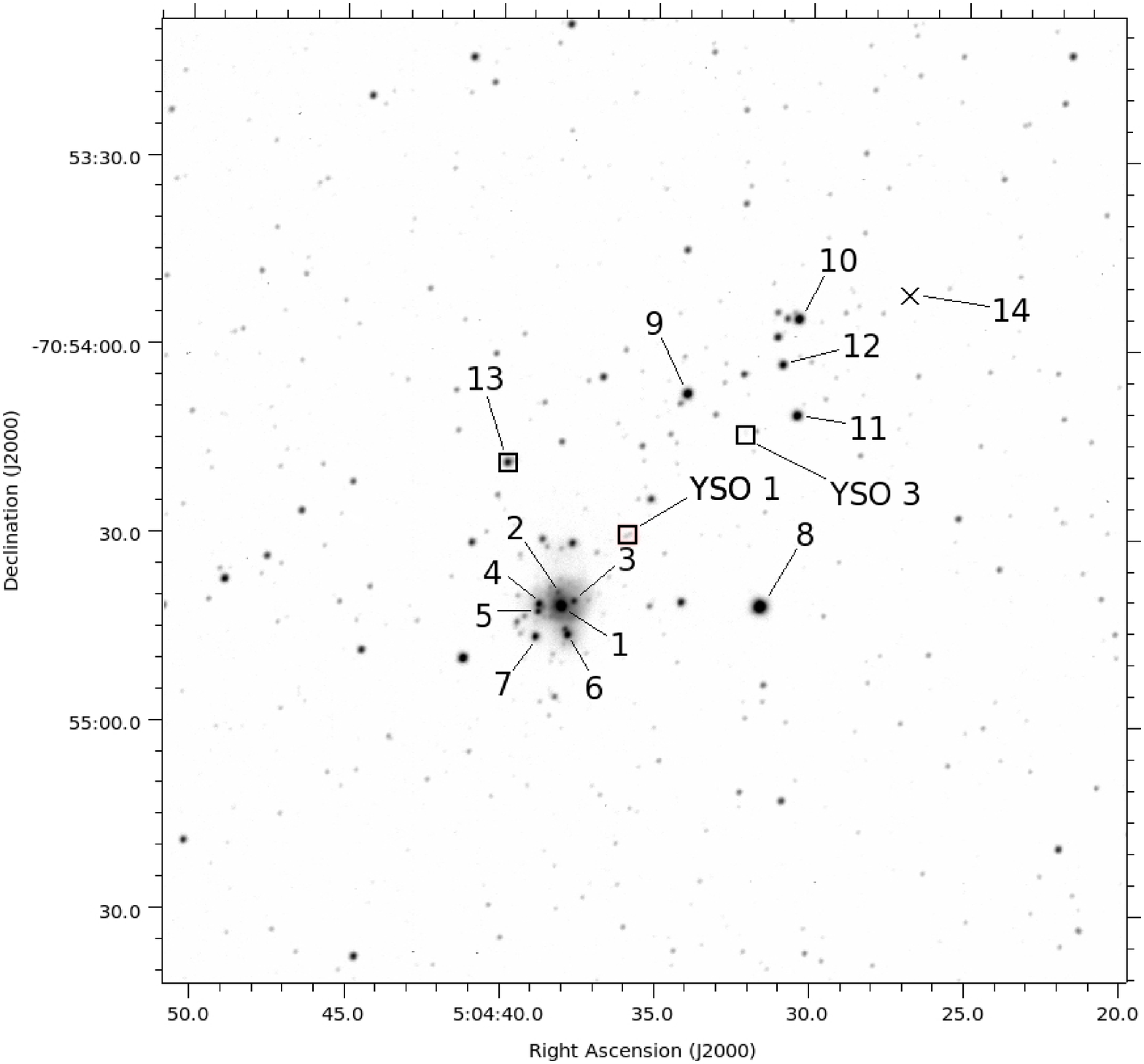} 
\caption{Large Magellanic Cloud \hii\ region N191. 
{\it Upper panel}: Composite three-color image showing the two components 
A and B. The main component, N191A, is situated at 
$\alpha$\,=\,05h\,04m\,38s and $\delta$\,=\,-70\deg\,\,54\min\,41\frac.  
This image, taken with the ESO NTT/SuSI2, is a coaddition of narrow-band 
filters \ha\ (red),  \oiii\ (green), and \hb\ (blue).  
The field size, 153\frac\ \x\,153\frac\, (\ab\ 37\,\x\,37 pc), is a  
close up of an original image covering a field of  319\frac\,\,\x\,327\frac\, 
(78\,\x\,79 pc). North is up and east to the left.  
{\it Lower panel}: The same field through the broad-band filter 
$V$. The brightest stars of the field and the young stellar object candidates 
are labeled (Tables \ref{tab:stars_n191} and \ref{tab:phot_yso}).
}
\label{fig:n191}
\end{figure*}

\begin{figure*}[]
\centering
\includegraphics[width=0.5\hsize]{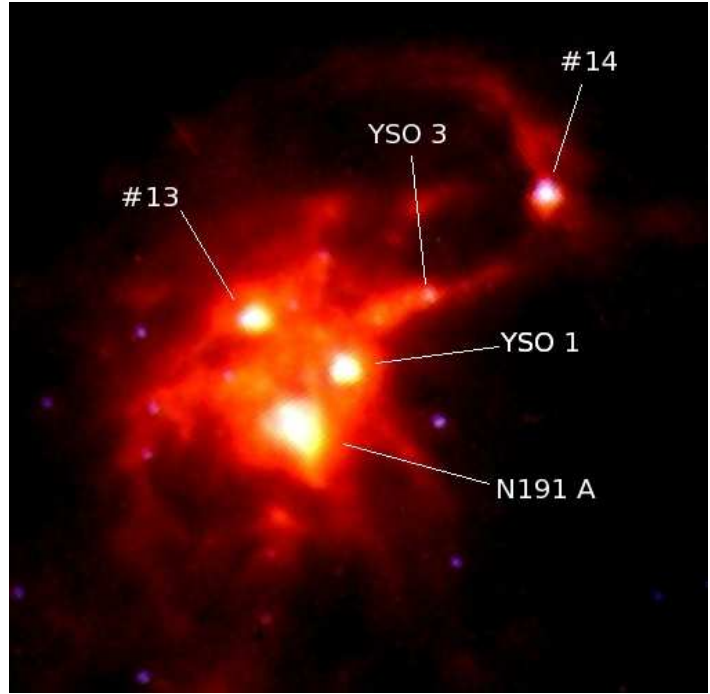}
\caption{Composite image of the LMC N191 region obtained with Spitzer IRAC. 
The 4.5 $\mu$m band is represented in blue, the 5.8 $\mu$m band in yellow, and 
the 8.0 $\mu$m band in red. The field size and orientation are the same 
as for Fig.\,\ref{fig:n191}. 
}
\label{fig:n191_spitzer}
\end{figure*}


\begin{figure*}[]
\centering
\includegraphics[width=0.7\hsize]{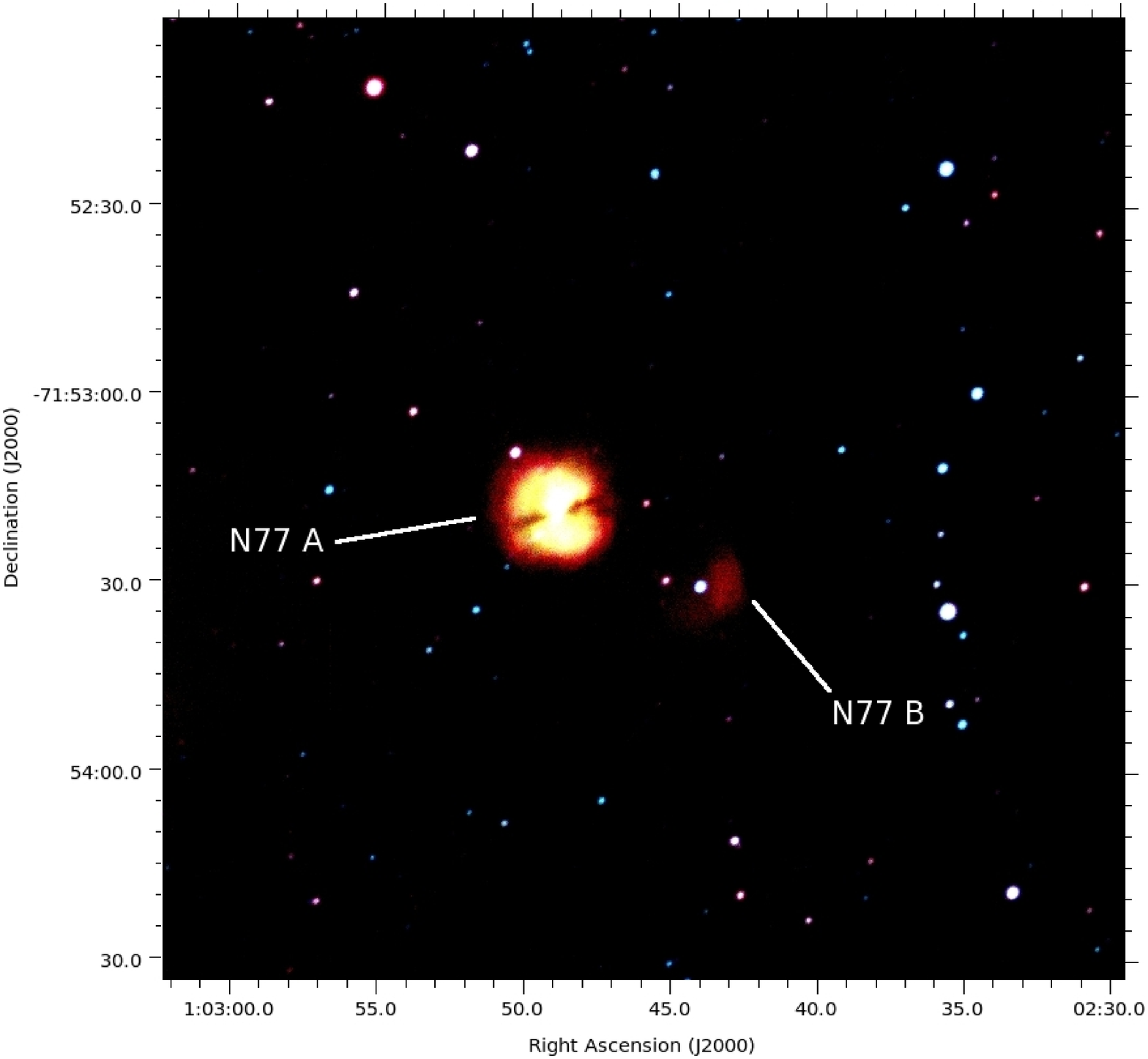}
\includegraphics[width=0.7\hsize]{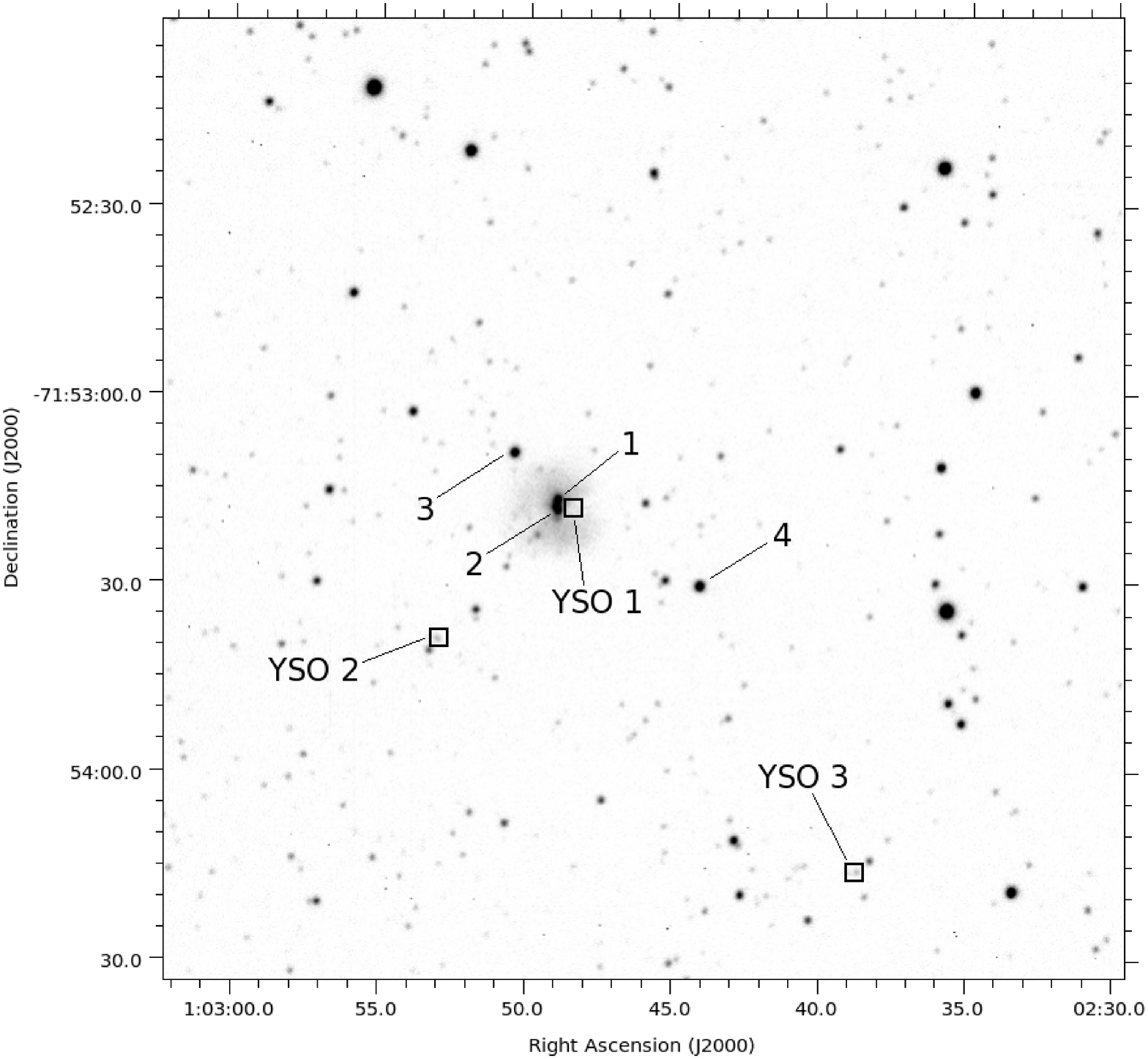}
\caption{
Small Magellanic Cloud \hii\ region N77. 
{\it Upper panel}: A composite three-color image showing the two components 
A and B. The main component, N77A, is situated at 
 $\alpha$\,=\,01h\,02m\,49s and $\delta$\,=\,-71\deg\,\,53\min\,18\frac.  
This image, taken with the ESO NTT/SuSI2, is the coaddition of narrow-band 
filters \ha\ (red),  \oiii\ (green), and \hb\ (blue).  
The field size 153\frac\ \x\,153\frac\, (\ab\ 45\,\x\,45 pc).
is a close up of an original image covering a field of 319\frac\,\,\x\,327\frac\, 
(93\,\x\,95 pc). North is up and east to the left.  
{\it Lower panel}: The same field through the broad-band filter 
$V$. The brightest stars of the field and the young stellar object candidates 
are labeled (Tables \ref{tab:stars_n191} and \ref{tab:phot_yso}).}
\label{fig:n77}
\end{figure*}

\begin{figure*}[]
\centering
\includegraphics[width=0.5\hsize]{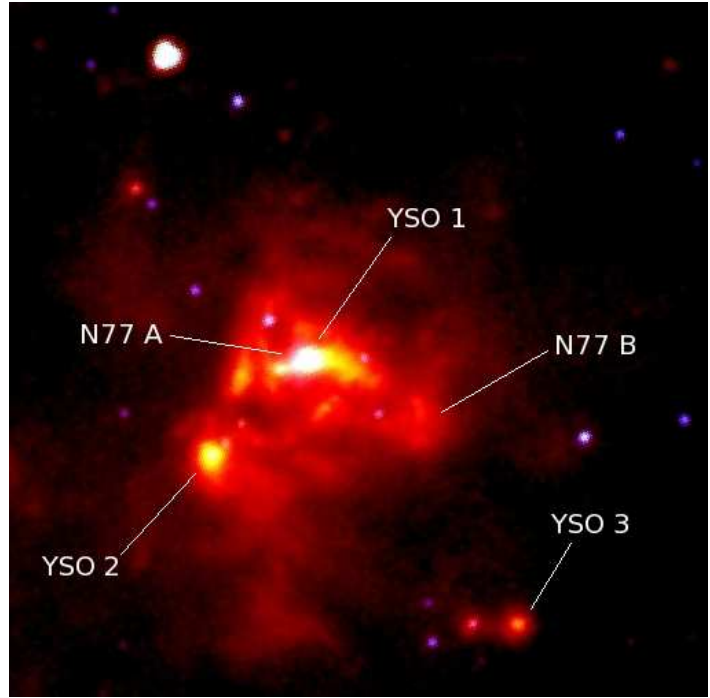}
\caption{Composite image of the SMC N77 region obtained with Spitzer IRAC. 
The 4.5 $\mu$m band is represented in blue, the 5.8 $\mu$m band in yellow, and 
the 8.0 $\mu$m band in red. The field size is the same as for Fig.\,\ref{fig:n77}. 
North is up and east to the left.}
\label{fig:n77_spitzer}
\end{figure*}

\section{Results}

\subsection{Overall view}

The images taken with the NTT telescope (Sect. 2.1) have a 
whole area of \ab\,5\min\,\x\,5\min\, corresponding to 
\ab\,73 pc\,\x\,73 pc for a distance of \ab\,50 kpc (LMC) or 
\ab\,90 pc\,\x\,90 pc for a distance of \ab\,60 kpc (SMC)  \citep[][]{Laney94}. \\ 

The LMC N191 \hii\  region \citep[][]{Henize56} consists of two components A and B in 
the optical, as displayed in Fig.\,\ref{fig:n191}. The brighter component A contains  
a compact \hii\ region on which we focus here. Component B  is comparatively very diffuse. 
The Spitzer observations show a richer nebulous region (Fig.\,\ref{fig:n191_spitzer}) 
compared to the optical, the brightest object of which is the compact \hii\ region N191A. 
The Spitzer observations 
also uncover two relatively bright, compact nebulae situated north of component A. 
These two objects correspond to star \#13 (Table\,\ref{tab:stars_n191}) and a 
young stellar object (YSO)
candidate  (Sect. 4). The component N191B is very weak or almost nonexistent in the Spitzer 
bands. It is crossed by a curl of gas emanating from the N191A region. 
The bright source, seen in the middle of the curl and  called \#14, 
has no noticeable optical counterpart.  
It is in fact an ``extreme AGB star'', as detected in the SAGE Survey and is cataloged 
as SSTISAGE1C J050426.95-705351.6 \citep[][]{Vijh09}. \\

The compact \hii\ region N191A has a 
mean angular radius, ($\theta_{\alpha}.\theta_{\delta})^{1/2}$, of 5\frac.2 corresponding 
to a radius of 1.2 pc. Broad-band images in $B$, $V$, and $R$ (Fig.\,\ref{fig:n191}, lower panel) 
reveal seven stars (\#1 to \#7) within less than 6\frac\ of the compact \hii\ region,   
whose positions and photometry are listed in Table\,\ref{tab:stars_n191}. 
We will  show that the central star \#1 is the exciting source of the \hii\ 
region (see Section 2.4).\\

The SMC N77 field presents a  different situation in the optical for the 
diffuse nebula. N77 is composed of two components 
A and B  (Fig.\,\ref{fig:n77}).
N77A contains two stars (\#1 and \#2) separated by only 1\frac.3.
This compact region of ionized gas appears as a sphere of radius \ab\ 10\frac\ (\ab\ 2.9 pc) split 
into two lobes by a dust lane that runs along an almost east-west direction. 
The two lobes intersect about 1\frac\ south of star \#2. 
Stars \#1 and \#2 are situated in the northern lobe, which 
is brighter in the Balmer lines \ha\ and \hb\ and also in the \oiii\ line. 
The \ha\ image shows faint emission around  star \#4 (Fig.\, \ref{fig:n77}) 
situated \ab\ 26\frac\, southwest of N77A. This nebula is known as N77B in the Henize catalog. 
The positions and photometry of stars \#1 and \#2 are listed in Table\,\ref{tab:stars_n191}. 
The Spitzer observations (Fig.\,\ref{fig:n77_spitzer}) display a curved structure focused 
on N77 in which the two lobes merge, but this is probably only because of 
low resolution. N77B is very 
dim in the Spitzer image. Fig.\,\ref{fig:n77_spitzer} also indicates three YSO 
candidates that will be discussed in Sect. 4.

\subsection{Extinction}

The map of the \ha\,/\,\hb\ Balmer decrement confirms that the \hii\ region LMC N191 is heavily 
affected by interstellar dust. The \ha\,/\,\hb\ ratio is on average  7.0 (A$_{V}$ = 2.7 mag) and  
up to 10.0 (A$_{V}$ = 3.8 mag) in the most extincted area.
The extinction toward star \#1 can be derived from a 
second method. O-type stars have an intrinsic color of {\it B\,--\,V}\,=\,-0.28 mag 
\citep[][]{Martins06}. This yields a color excess of {\it E(B\,--\,V)}\,=\,0.39 mag 
or a visual extinction of A$_{V}$ = 1.2 mag. This value 
is lower than the result from the Balmer decrement because  
the central regions of the \hii\ region are less affected by extinction. 
The dust has been more dispersed along the line of sight of star \#1. Moreover, the extinction 
toward LMC N191 was estimated  
by a third method using radio continuum observations. N191 appears as the 
source B0505-7058 in the Parkes radio continuum survey at 2.45, 4.75 and 8.55 GHz, which had 
beam-sizes of 
8\min.85, 4\min.8 and 2\min.7, respectively \citep[][]{Filipovic95}.
The resulting extinction, A$_{V}$ = 2.4 mag, is comparable with that obtained 
using the previously mentioned methods in the optical range.\\

The average value of the Balmer decrement toward the \hii\ region SMC N77 is 
about 3.1, corresponding to A$_{V}$ = 0.25 mag.
The most extincted part of the \hii\ region is its western border, along the dust lane 
(see Section 3.1), where the \ha\,/\,\hb\ ratio reaches a value of 4.5 (A$_{V}$ = 1.4 mag). 
We also derived the extinction from the radio observations of SMC N77. 
High-resolution observations of this object were obtained by \citet[][]{Filipovic02}, who 
used the Australia Telescope Compact  
Array (ATCA) in radio continuum emission at 1.42, 2.37, 4.80 and 8.64 GHz with 
synthesized beams of 98\frac , 40\frac , 30\frac\ and 15\frac , respectively. 
The resulting extinction, A$_{V}$ = 0.56 mag, is comparable with that inferred 
using the Balmer decrement.

\subsection{Nebular emission}

The total H$\beta$ fluxes of the compact \hii\ regions LMC N191 and SMC N77 were derived 
using the following method. First we calculated the relative H$\beta$ flux in an 
imaginary 1\frac\, slit passing through the H$\beta$ image with respect to 
the total flux emitted by the whole \hii\ region. This value was then 
compared with the absolute flux obtained from the spectra. The total H$\beta$ flux thus obtained for N191 was
$F$(\hb )\,=\,1.52\,\x\,10$^{-12}$ erg cm$^{-2}$ s$^{-1}$. 
Studies of the extinction in the LMC and the SMC reveal 
reddening laws that are similar to the average Galactic law for the optical 
and near-IR wavelengths \citep[][]{Howarth83,Prevot84,Bouchet85}.
The reddening coefficient, derived from the mean  \ha\,/\,\hb\ ratio of 7, 
was {\it c}(\hb\,) = 1.26. Considering the extinction law for the LMC \citep{Howarth83}, 
we computed the reddening corrected intensity 
$I$(\hb )\,=\,2.77\,\x\,10$^{-11}$ erg cm$^{-2}$ s$^{-1}$. \\

This flux corresponds to a Lyman continuum flux of 1.78\,\x\,10$^{49}$ 
photons s$^{-1}$ for the star, assuming that the \hii\ region is ionization-bounded. 
The exciting star needed to provide this flux should have an effective temperature of 
\ab\,41000 K, corresponding to a spectral type about O5\,V, 
for Galactic metallicity \citep[][]{Martins05}. 
However, this flux is probably underestimated since the \hii\ region is likely not
completely ionization-bounded. \\

Similarly, the total H$\beta$ flux obtained for N77 was
$F$(\hb )\,=\,2.03\,\x\,10$^{-12}$ erg cm$^{-2}$ s$^{-1}$. 
Considering the extinction law for the LMC \citep{Howarth83}, the corrected flux was 
$I$(\hb )\,=\,2.61\,\x\,10$^{-12}$ erg cm$^{-2}$ s$^{-1}$ 
with the reddening coefficient {\it c}(\hb\,) = 0.11.
This flux value corresponds to a Lyman continuum flux of 2.4\,\x\,10$^{48}$ 
photons s$^{-1}$ for the star. 
The exciting star needed is of spectral type about O8\,V, 
for Galactic metallicity \citep[][]{Martins05}. 
This may be a lower limit, however, because of photon loss in a density-bounded \hii\ region.\\

Several of the derived physical parameters of the compact \hii\ regions are summarized in 
Table\,\ref{tab:param}. The mean angular radius of the \hii\ region, corresponding to 
the FWHM of cross-cuts through the \ha\ image, is given in Col. 2. The corresponding 
physical radius, obtained using distance moduli of {\it m\,-\,M} = 18.53 mag for LMC N191 
and {\it m\,-\,M} = 18.94 mag for SMC N77 \citep[][]{Laney94} is presented 
in Col. 3. The reddening coefficient, derived from the mean  \ha\,/\,\hb\ ratio, 
is listed in Col. 4. It corresponds to the whole \hii\ region. It is different from the value 
found from the nebular spectrum (Table\,\ref{tab:flux})   
because, in contrast, the spectrum belongs to a particular position and therefore 
does not cover the whole region. 
The de-reddened \hb\ flux obtained from the reddening coefficient is given in Col. 5 
and the corresponding \hb\ luminosity in Col.6. The electron temperature is given in Col. 7.
For N77A the electron temperature is calculated from the forbidden-line ratio 
\oiii\,\lam \lam\,4363/(4959 + 5007), with an uncertainty of 4\%. For N191A, 
the \oiii\,\lam\,4363 is not observed in our spectra 
so we used the electron temperature calculated from the forbidden-line ratio 
\oii\,\lam \lam\,(3726 + 3729)/(7323 + 7330), with an uncertainty of 10\%, higher than the 
estimate from the \oiii\ ratio.
The electron density, estimated from the ratio of the \sii\ doublet  \lam \lam\,6717/6731, 
is presented in Col. 8. It is accurate to \ab\,80\%. 
It is well-known that the  \sii\ lines characterize the low-density 
peripheral zones of \hii\ regions. 
Column 7 gives the rms electron density, {\it $<$n$_{e}$$>$}, 
calculated from the \hb\ flux, the radius, and 
the electron temperature, {\it T$_{e}$}, assuming that the \hii\ region is 
an ionization-bounded Str\"omgren sphere. Furthermore,  
the total mass of the ionized gas, calculated from the {\it $<$n$_{e}$$>$} 
with the previously noted Str\"omgren sphere assumption is presented in Col. 9. The ionization 
is produced by Lyman continuum photon flux given in Col. 10.

\begin{table*}
\caption{Some physical parameters of the compact \hii\ regions SMC N77 and LMC N191}
\label{tab:param}  
\begin{tabular}{l c c c c c c c c c c} 
\hline\hline
Object & $\theta$ & $r$  & $c$(\hb )    &   $I$(\hb ) &  $L$(\hb ) & {\it Te}    & {\it Ne}$^{\dag}$   & {\it $<$n$_{e}$$>$}  
& {\it M$_{gas}$}  & $N_{L}$ \\
& (\frac\,) & (pc) &  & erg s$^{-1}$ cm$^{-2}$  &  erg s$^{-1}$  & (K) & cm$^{-3}$ & cm$^{-3}$ & (\sm ) & ph s$^{-1}$ \\
   &  &      &    & \x\,10$^{-12}$ &   \x\,10$^{36}$  &   &      &      &        &  \x\,10$^{48}$ \\
\hline
  LMC N191A &  5.2  &    1.2  &  1.26 & 27.7 & 8.3 &  10800  & 440     &  600    &   140  &  17.8 \\
  SMC N77A  & 10    &    2.9  &  0.11 &  2.6 & 1.1 &  14240  &  40     &   60    &   200  &   2.4 \\
\hline

\end{tabular} \\

$\dag$ Estimated from the \sii\ ratio.
\end{table*}

\subsection{Stellar content} 

\subsubsection{LMC N191}

The images show some 15 relatively bright stars lying within 10\frac\, of N191A. 
The brightest component of this group, star \#1, has $V$ = 14.46 mag and it is 
followed by stars \#6 and \#4 with $V$ = 16.28 and 16.60 mag, respectively. 
Assuming an intrinsic color of {\it B\,--\,V}\,=\,-0.28 mag for O-type stars  
\citep[][]{Martins06} and a distance modulus of 18.53 mag, the absolute magnitude 
of star \#1 is M$_{V}$ = -5.27 mag. Following the calibration of \citet[][]{Martins05} for 
Galactic stars, if the star is on the main sequence, it would be of a spectral type O5\,V with 
a mass of 40 \sm . This agrees well with what was found based on the stellar 
Lyman continuum derived from the \hb\, emission of the \hii\ region (see Sect. 3.3). \\

\begin{figure*}[]
  \centering
\includegraphics[width=14cm]{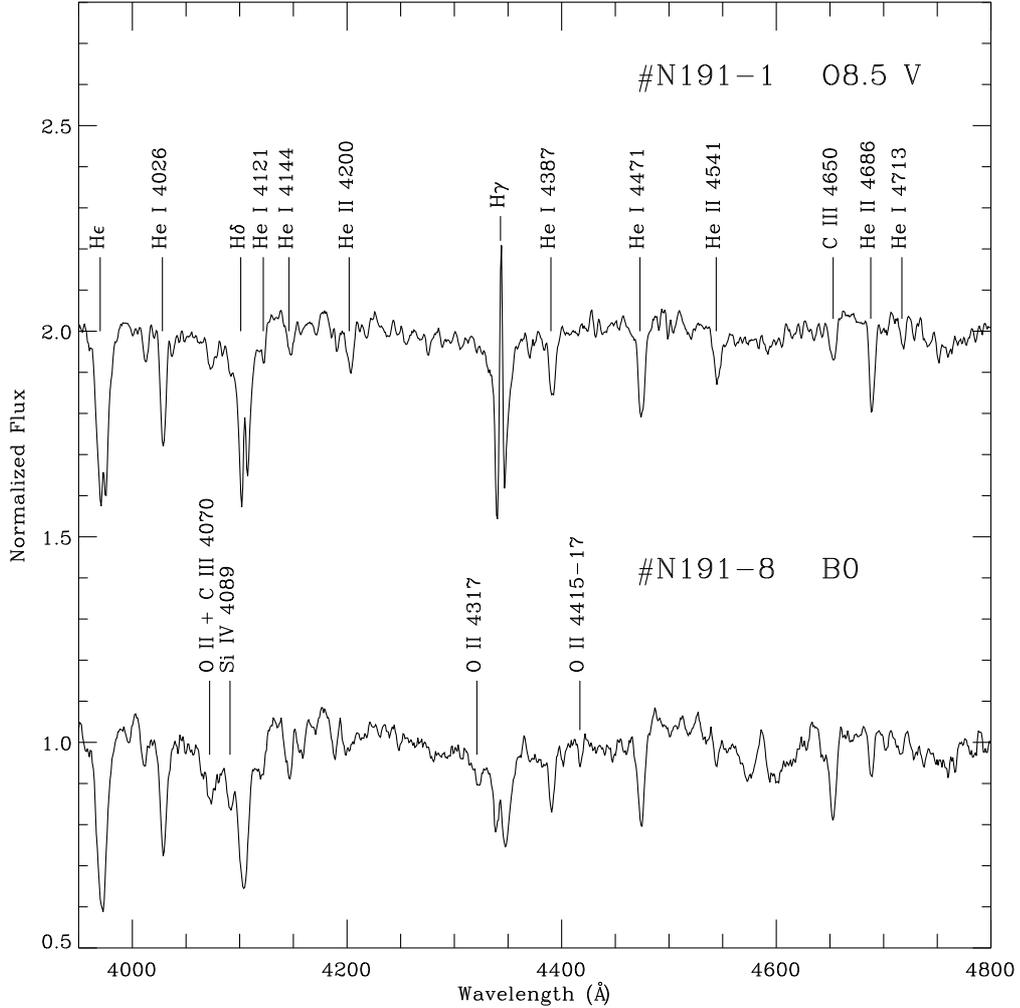}
\caption{
Spectra of two stars observed toward LMC N191. Star \#1 
is the exciting source of the compact \hii\ region N191A. Note the  
\heii\ absorption lines indicating a hot massive star O8.5\,V. 
Star \#N191-8, lying in the field of N191, is classified 
as B0.}  
\label{fig:spectres_n191}
\end{figure*}

We obtained spectra of two stars of the N191 region in our program of stellar spectroscopy 
(Sect. 2.2), in particular star \#1, for the first time. 
However, in spite of the relatively good seeing conditions, 
extracting uncontaminated spectra is not straightforward. The compact 
\hii\ region has strong emission lines, in particular those of \hei , that fill in the 
absorption lines of the embedded stars. Nevertheless, we classified \#N191-1 as O8.5\,V with 
a good accuracy (Fig.\,\ref{fig:spectres_n191}). The spectral classification was performed 
using the criteria stated by \citet[][]{Walborn90}.
This is colder than indirect estimates and more uncertain
(absolute magnitude and stellar Lyman continuum). However, we cannot exclude that 
the spectrum of N191A-1 is not contaminated by unresolved close companions of later type than 
star \#1. See Sect. 4 for discussion. \\
 
The spectrum of star \#N191-8, the brightest object of the field, with $V$ = 13.79 mag, 
lying \ab\,30\frac\ west of star \#1, outside N191A, is also presented in 
Fig.\,\ref{fig:spectres_n191}. 
Using the same classification criteria \citep[][]{Walborn90}, we can assign spectral 
type B0 to this star. The spectrum does not allow one  
to conclude firmly on the luminosity class 
because the B-type luminosity criteria are not sufficiently apparent here. 
Taking into account an extinction of $A_{V}$ = 1.2 mag, the absolute magnitude of this 
star is M$_{V}$ = -5.98 mag. This determination points to a supergiant 
Ib \citep{Fitzpatrick90}. Therefore star \#N191-8 likely provides a part of the ionizing 
photons that power the compact \hii\ region. See below for more details about 
the evolutionary stage of this star. \\

\begin{figure*}[]
\centering
\includegraphics[width=1.0\hsize]{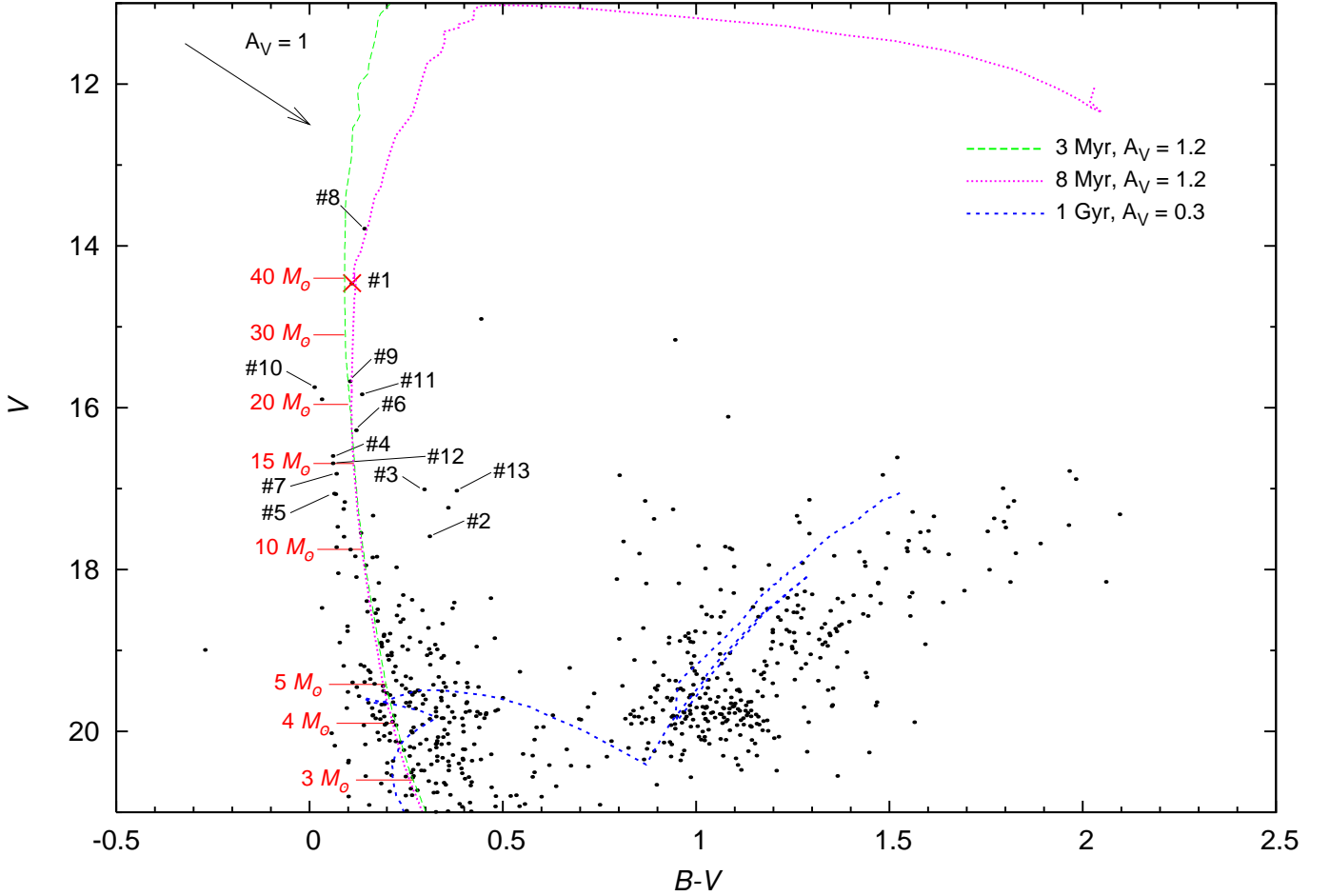}
\caption{Color-magnitude {\it V} versus {\it B -- V} diagram for 
stars observed toward LMC N191. 
Three  isochrones are shown, 3 Myr ($A_V = 0.9$ mag, 
green dashed curve), 8 Myr ($A_V = 1.2$ mag, violet dotted curve) and 
1 Gyr ($A_V = 0.3$ mag, blue thick dashed curve), computed for a
metallicity of Z = 0.008 \citep{Lejeune01} and a distance modulus of 18.53 mag.  
The red cross indicates the location of the main exciting star of N191A. 
The numbers refer to the stars listed in Table\,\ref{tab:stars_n191}.
}
\label{fig:cmd_n191}
\end{figure*}

The color-magnitude diagram of the star population 
in the entire NTT field of 319\frac\,\,\x\,327\frac\ (\ab\,78\,\x\,79 pc) 
for a cut-off magnitude of $V$ = 21 is presented in Fig. \ref{fig:cmd_n191}. 
Three isochrones with ages 3 Myr, 8 Myr, and 1 Gyr, for a metallicity of 
Z = 0.008 \citep{Lejeune01}, are also overplotted. 
The diagram displays two principal groups: an apparent main sequence centered on 
$B-V$\,\ab\,0.1  mag and an evolved population centered on $B-V$\,\ab\,1 mag.  \\

Star \#1, indicated by a cross, is affected by an extinction of A$_{V}$ = 1.2 mag 
therefore it seems reddened compared to the 3 Myr isochrone. 
Accordingly, star \#1 appears to be the most 
massive young star of the field. The stars lying across the \hii\ region N191 are concentrated 
along the main sequence. They are intermediate-mass stars of \ab\ 10-15 \sm\ assuming 
that they are on the 3 Myr isochrone. \\

Star \#8, which we described above, is apparently  
related to the 8 Myr isochrone. This means that star  \#8 is probably older than 
the exciting star of the compact \hii\ region. The fact that no noticeable 
surrounding ionized region is associated with star \#8 is compatible with this deduction. 
An initial mass of 22 \sm\ can be derived for this evolved B-type star. \\

The second group of stars on the diagram are evolved stars between {\it B\,--\,V}\,=\,0.8 
and 1.6 mag. They have a possible age ranging from 1 Gyr to 10 Gyr.  It is very likely 
that this latter population is not physically associated with N191.

\subsubsection{SMC N77A}

The stellar environment of N77 is totally different from that of N191.
The images reveal only two stars embedded in the blob.
These two stars have similar visual magnitudes ($V$ = 17.35 and 17.57 mag), and are 
fainter than the main N191 stars. 
Assuming that O-type stars have an intrinsic color of {\it B\,--\,V}\,=\,-0.28 mag 
\citep[][]{Martins06} (A$_{V}$ = 2.1 mag) and a distance modulus of 18.94 mag, 
the absolute magnitude of star \#1 is M$_{V}$ = -3.69 mag. It is too faint to be an 
O star \citep[][]{Martins05} unless the extinction is underestimated. \\

\begin{figure*}[]
  \centering
\includegraphics[width=14cm]{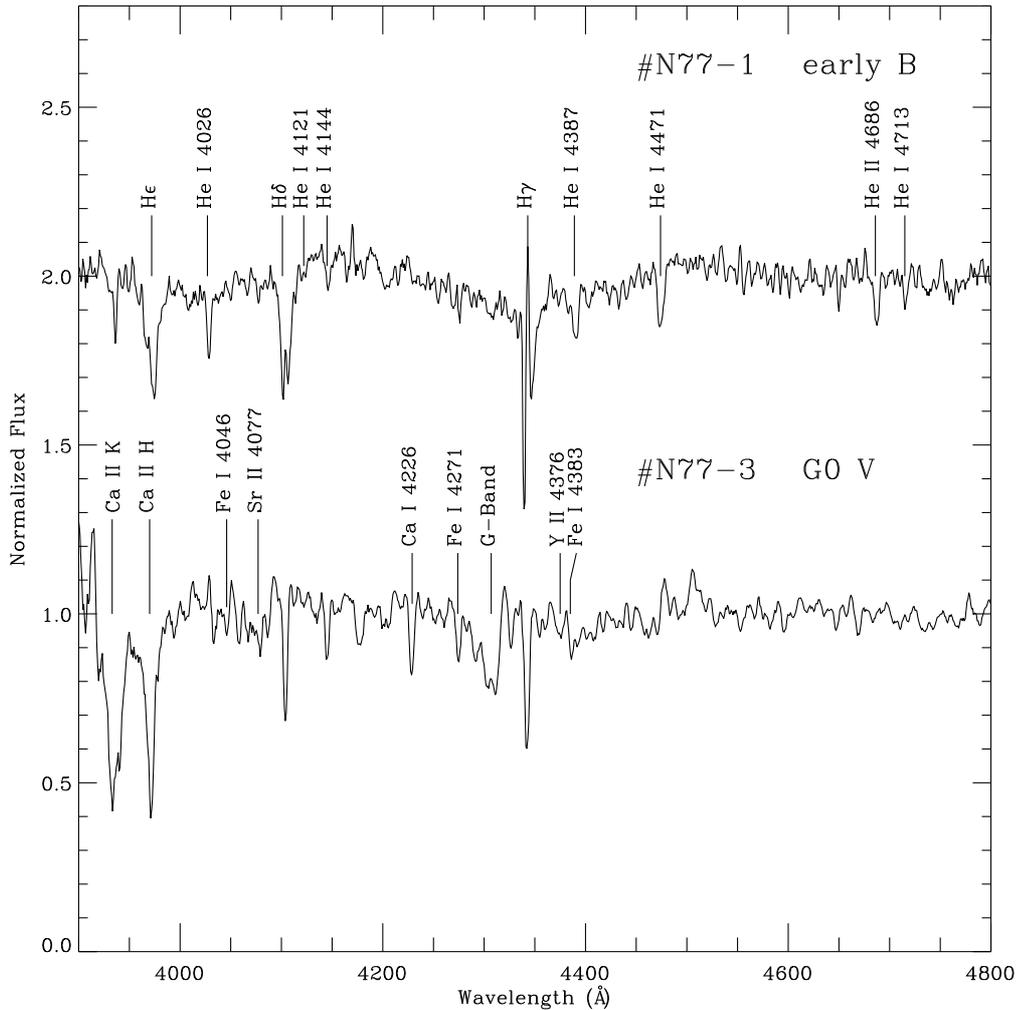}
\caption{
Spectrograms of two stars observed toward SMC N77A. Star \#1, an early B-type star, 
is the exciting source of the compact \hii\ region N77A.
Star \#3 is classified as a Galactic star G0\,V.} 
\label{fig:spectres_n77}
\end{figure*}

Fig.\,\ref{fig:spectres_n77} displays the spectrum of star \#1, or more precisely the 
spectrum of stars \#1 and \#2. Indeed, the two stars are too close to obtain separate spectra. 
Nevertheless, the \heii\ absorption line at \lam\ 4686 is 
certainly present, while in contrast \heii\ \lam\ 4541 is not observed. 
These features indicate that N77-1 is an early-type B star. This agrees 
with the estimate of the absolute magnitude. However, it is significantly colder than 
the spectral type inferred from the \hb\ flux measurement 
(see Sect. 3.3 and Sect. 4).

\begin{table*}             
\caption{Positions and photometry of the main stars in the fields of LMC N191 and SMC N77 $^{\dag}$}  
\label{tab:stars_n191} 
\begin{tabular}{l l c c c c c c c r c c}     
\hline\hline
Galaxy & star ID & $\alpha$ (J2000) & $\delta$ (J2000) & $V$ & $B-V$ & $V-R$ & $J$ & $H$ & $K$ 
& spectral type & \hii\ region \\
& & & &  & & & & & & & component\\
\hline
LMC & N191-1  & 05:04:38.12 & -70:54:41.28 & 14.46 &  0.11 &  -0.01 & 13.29 & 13.33 & 12.05 & O8.5 V & A \\
    & N191-2  & 05:04:38.23 & -70:54:39.05 & 17.59 &  0.31 &  0.43 &       &       &        &        & A\\
    & N191-3  & 05:04:37.72 & -70:54:40.50 & 17.01 &  0.30 &  0.19 &       &       &        &        & A\\
    & N191-4  & 05:04:38.84 & -70:54:40.95 & 16.60 &  0.06 &  0.14 &       &       &        &        & A\\
    & N191-5  & 05:04:38.86 & -70:54:42.11 & 17.06 &  0.06 &  0.17 &       &       &        &        & A\\
    & N191-6  & 05:04:37.92 & -70:54:45.78 & 16.28 &  0.12 &  0.16 & 14.37 & 14.11 & 12.36  &        & A\\
    & N191-7  & 05:04:38.96 & -70:54:46.16 & 16.82 &  0.07 &  -0.08 &      &       &        &        & A\\
    & N191-8 & 05:04:31.71 & -70:54:41.06 & 13.79 &  0.14 &  0.04 & 13.16 & 13.06 & 12.91 & B0      & A\\
    & N191-9  & 05:04:34.08 & -70:54:07.21 & 15.67 &  0.10 &  -0.04 & 15.35 & 15.15  & 15.02 &      & B \\
    & N191-10  & 05:04:30.49 &  -70:53:55.11 & 15.74 &  0.01 &  -0.11 & 15.53 & 15.78  & 15.28 &    & B\\
    & N191-11 & 05:04:30.53 & -70:54:10.51 & 15.83 &  0.13 &  0.03 & 15.32 & 15.18  & 14.88 &       & B\\
    & N191-12  & 05:04:31.00 & -70:54:02.42 & 16.69 &  0.06 &  -0.11 &  &   &  &                    & B\\
    & N191-13 & 05:04:39.85 & -70:54:19.00 & 17.02 & 0.38 & 0.38 & 15.00 & 14.45 & 13.71 &          &  \\
    & N191-14 & 05:04:26.97 & -70:53:51.71  &      &      &     &   16.34 & 14.01 & 12.44 &  AGB star        &  \\
SMC & N77-1  & 01:02:48.98 & -71:53:16.58 & 17.35 &  0.40 &  -0.23 & 14.75 & 14.51 & 14.03 & early B & A \\
    & N77-2  & 01:02:49.04 & -71:53:17.69 & 17.57 &  0.30 &  0.23 &       &       &       &          & A \\
    & N77-3  & 01:02:50.46 & -71:53:09.10 & 17.01 &  1.53 &  0.23 &  14.13   &  13.51  &  13.45  & G0 V (Galactic) & \\
    & N77-4  & 01:02:44.12 & -71:53:30.30 & 17.06 &  0.29 &  -0.38 & 16.38 & 16.02 & 14.98 &          & B\\
\hline
\end{tabular} \\

$\dag$
The {\it BVR} photometry results from the NTT observations while the 
{\it JHK} measures are taken from the 2MASS catalog. \\
\end{table*}

\begin{table*}             
\caption{Positions and photometry of the YSO candidates in the fields of LMC N191 and SMC N77 $^{\dag}$}  
\label{tab:phot_yso} 

\begin{tabular}{l l l c c c c c c c c c c}     
\hline\hline
Field & YSO ID & star ID & $\alpha$ (J2000) & $\delta$ (J2000) & $J$ & $H$ & $K$ & [3.6] & [4.5] & [5.8] & [8.0]  & [24.0] \\
\hline
LMC N191 & YSO 1  &  & 05:04:35.85 & -70:54:30.1 & 15.75 & 14.73 & 13.33 &  10.48 & 9.54 &  8.09  &  6.56  & 1.83\\
    & YSO 2 & N191-13 & 05:04:39.85 & -70:54:19.0 & 15.00 & 14.45 & 13.71 &  11.28  &  10.57  & 8.68  & 6.95 & 0.85 \\
    & YSO 3  & & 05:04:32.19 & -70:54:14.0 &      &       &       &  13.32   &  12.73 & 10.95 & 9.34 & 4.84\\
SMC N77 & YSO 1  & & 01:02:48.54 & -71:53:18.0 &      &        &    & 13.04 & 12.09 & 10.42 &  8.93   & 3.55\\
    & YSO 2  & & 01:02:53.13 & -71:53:39.2 & 16.42 & 16.02 & 15.51   &  13.80 & 13.89 &  10.87 & 9.19  & 4.59\\
    & YSO 3  & & 01:02:38.81 & -71:54:15.9 &      &       &      &  14.89   &      &  12.83  & 11.17 & 8.46 \\
\hline
\end{tabular} \\

$\dag$
The {\it JHK} measures are taken from the 2MASS catalog. 
IRAC and MIPS photometry come from \citet{Bolatto07} for the SMC field and from \citet{Gruendl09} for the LMC field. 

\end{table*}

\begin{figure*}[]
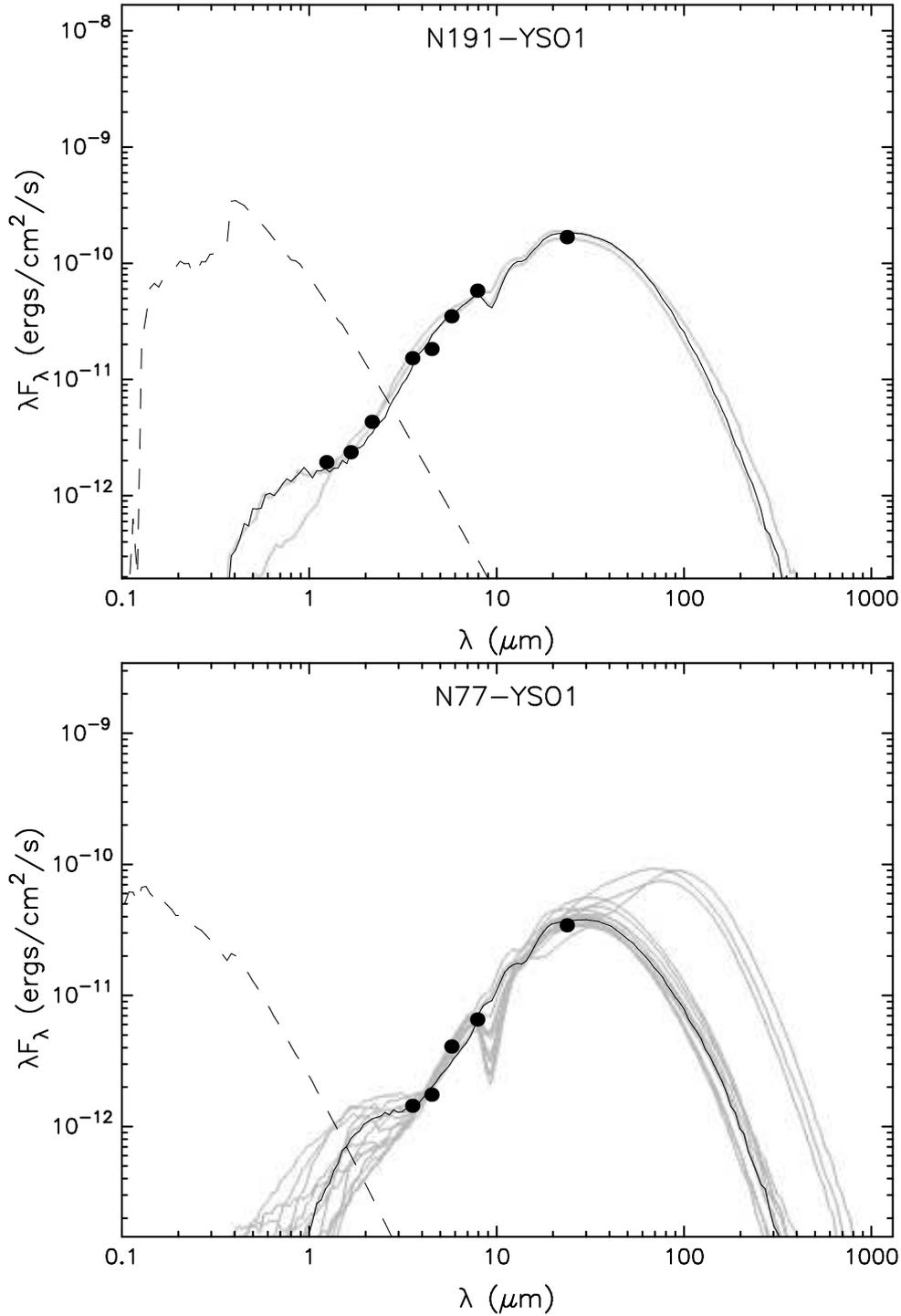

\centering

\includegraphics[width=0.7\hsize]{19706fg8a.eps}
\includegraphics[width=0.7\hsize]{19706fg8b.eps} 
\caption{
Mid-IR Spitzer photometry of two YSOs fitted using 
YSO models \citep{Robitaille07}.  
The filled circles represent the input fluxes. The black curve shows the best fit, 
while the gray curves display subsequent good fits. The dashed curve shows the stellar 
photosphere corresponding to the central source of the best-fitting model, as it 
would look in the absence of circumstellar dust (but including interstellar extinction). 
{\it Upper panel}: SED fit of N191-YSO1. The stellar mass according to these models 
varies between 16 and 21 \sm . 
The best fit is for a 20 \sm\ protostar with a derived total luminosity of 2.9\,\x\,10$^{4}$ \slum .
{\it Lower panel}: SED fit of N77-YSO1. The stellar mass according to these models varies 
between 10 and 17 \sm . Note that the 
lack of $JHK$ photometry leads to a wider range of selected models.
The best fit is for a 10 \sm\ protostar with a derived total luminosity of 
1.0\,\x\,10$^{4}$ \slum . 
}
\label{fig:yso}
\end{figure*}

\subsection{Chemical abundances}

Table\,\ref{tab:flux} lists the main lines of the nebular spectra of N191 and N77. 
The ionic abundances with respect to H$^{+}$  were calculated from nebular lines 
using the IRAF task ionic of the package NEBULAR \citep{Shaw95}. The abundance values are 
listed  in Table\,\ref{tab:ion}. \\

To derive the total abundances of a given element, it is necessary to estimate the amount of 
the  element in ionization states not observed in our spectra.
We therefore used a set of ionization-correction factors (ICFs) to convert into 
elemental abundances. 
The absence of the nebular \heii\ line indicates that He$^{2+}$/H$^{+}$ is negligible. 
Moreover, we assume that neutral helium is not important. Thus we 
assumed that the total He/H ratio is just equal to He$^{+}$/H$^{+}$.
The total abundance of oxygen was adopted to be the sum of the O$^{+}$ and O$^{2+}$ abundances.
The absence of \heii\ recombination lines in our spectra and the similarity between 
the ionization potentials of He$^+$ and O$^{++}$ implies that the contribution of O$^{3+}$ 
is not significant.
To obtain the total abundance of nitrogen, we used the usual ICF
based on the similarity between the ionization potential of 
N$^{+}$ and O$^{+}$ \citep{Peimbert69}. The N$^{+}$ abundance does not depend strongly 
on the electron temperature. The largest errors come from the uncertainty in the 
\lam \lam\,6548 and 6584 line intensities. Our N result is accurate to within about 
30\%. The only measurable lines in the optical range for Ne are those of 
Ne$^{2+}$ but the amount of Ne$^{+}$ may be significant in the region. 
We adopted the usual expression of the ionization correction factor of Ne  
that assumes that the ionization structure of Ne is similar to that of O \citep{Peimbert69}. \\

The total chemical abundances for N191 and N77 are presented in Table\,\ref{tab:ab}.
The most accurately estimated abundances belong to He and O, which are accurate to within 
15\% and 20\%, respectively. Table\,\ref{tab:ab} also presents the mean abundance 
values derived for the SMC and the LMC \citep[][]{Russell92}. The N77 abundances agree well with 
the SMC mean values.
For N191, the He abundance is quite low compared with the LMC mean value. The strong stellar \hei\ absorption lines 
could have contaminated the nebular spectra of the \hii\ region.
The abundance of O, N and Ne are lower than the LMC mean value but this can be explained by 
uncertainties and also by the fact that N191 is more metal-poor than the LMC because of its 
external position.

\begin{table*}
\caption{Nebular line intensities of the compact \hii\ region LMC N191A and SMC N77A }
\label{tab:flux}
\begin{tabular}{llcccccc}
\hline  \hline	
  \multicolumn{2}{c}{} &  \multicolumn{3}{c}{LMC N191A} & \multicolumn{3}{c}{SMC N77A}\\
		$\lambda$ (\AA) & Iden. & $F$(\lam\,)$^{\dag}$ & $I$(\lam\,)$^{\dag}$
                 & Accuracy$^{\ddag}$ & $F$(\lam\,)$^{\dag}$ & $I$(\lam\,)$^{\dag}$ 
                 & Accuracy$^{\ddag}$\\
  \hline	
	3727,29 & \oii\           & 279.8  &  344.0 & A & 189.8 & 229.4 & A\\
	3797    & H10	          &  2.7  &   3.3 & B & 3.9 & 4.6 & B\\
	3835    & H9              &  3.6  &   4.3 & B & 5.6 & 6.6 & B\\
	3869    & \neiii\         &  2.1  &   2.5 & C & 12.7 & 14.9 & A\\
	3889,90 & \hei\ + H8      &  9.9  &  11.7 & A & 13.6 & 15.9 & A\\
	3968,70 & \neiii\ + \he\  &  9.7  &  11.3 & A & 14.5 & 16.7 & A\\
	4071    & \sii\           &  1.2  &   1.4 & D & & \\
	4101    & \hd\            & 39.1  &  44.3 & A & 19.0 & 21.3 & A\\
	4144    & \hei\           &  1.3  &   1.5 & D & & \\
	4340    & \hg\            & 34.6  &  37.5 & A & 40.3 & 43.4 & A\\
	4363    & \oiii\          &       &       &   & 4.8 & 5.1 & B\\
	4471    & \hei\           &       &       &   & 4.3 & 4.5 & C\\
	4861    & \hb\            & 100.0 & 100.0 & A & 100 & 100 & A\\
	4959    & \oiii\          & 49.7 & 49.1 & A & 104.6 & 103.4 & A\\
	5007    & \oiii\          & 147.4 & 144.6 & A & 300.2 & 295.0 & A\\
	5577    & \oi\            &   20.2 &   18.6 & A & 45.2 & 41.9 & A\\
	5876    & \hei\           &  10.5 &  9.4 & B & 11.1 & 10.0 & B\\
	6300    & \oi\            &   9.8 &   8.5 & C & 73.8 & 64.5 & A \\
	6312    & \siii\          &   1.2 &   1.0 & C & 2.0 & 1.7 & C \\
	6363    & \oi\            &   3.1 &   2.7 & D & 23.4 & 20.4 & A\\
	6548    & \nii\           &   13.2 &   11.2 & C & 10.1 & 8.7 & C\\
	6563    & \ha\            & 340.0 & 286.0 & A & 335.7 & 286 & A\\
	6584    & \nii\           &  37.9 &   32.0 & B & 11.6 & 9.9 & B\\
	6678    & \hei\           &   2.7 &   2.3 & B & 3.2 & 2.7 & B\\
	6716    & \sii\           &  17.2 &   14.4 & B & 15.8 & 13.4 & B\\
	6731    & \sii\           &   15.9 &   13.3 & B & 11.6 & 9.9 & B\\
	7065    & \hei\           &   2.4 &   2.0 & C & 2.8 & 2.3 & C \\
	7135    & \ariii          &  8.7 &   7.1 & B & 9.2 & 7.6 & B \\
	7236    & \ariv\          &   1.9 &   1.5 & D & 14.2 & 11.7 & B\\
	7323    & \oii\           &  12.9 &   10.3 & C & 22.0 & 18.0 & C \\
	7751    & \ariii\         &    3.9 &  3.0  & C & 16.2 & 12.9 & B\\
 \hline
c(H$\beta$) =                &                 &  0.24 &   &    &   0.22 & & \\     
   \hline 
\end{tabular}

$\dag$ $F(\lambda)$ and $I(\lambda)$ represent 
observed and de-reddened line intensities relative to \hb . \\
$\ddag$ The capital letters represent the following  uncertainties: 
A $<$\,10\%, B=10--20\%, C=20--30\%, and D$>$\,30\%.
\end{table*}

\begin{table*}
\caption{Nebular ionic abundances}
\label{tab:ion}
\begin{tabular}{lcc}
\hline\hline
Ion        & LMC N191 & SMC N77   \\
\hline
He$^+$/H$^+$                   & 0.070 & 0.079   \\
O$^+$/H$^+$ (\x\,$10^{5}$)     & 14.9 & 6.76 \\
O$^{++}$/H$^+$ (\x\,$10^{5}$)  & 5.10 & 3.63 \\
N$^+$/H$^+$ (\x\,$10^{6}$)     & 6.4 & 2.5  \\
Ne$^{++}$/H$^+$ (\x\,$10^{6}$) & 2.77 & 4.75 \\
S$^+$/H$^+$ (\x\,$10^{7}$)     & 6.86 & 5.33 \\
Ar$^{++}$/H$^+$ (\x\,$10^{7}$) & 6.54 & 3.36 \\
\hline	
\end{tabular}
\end{table*}

\vspace{0.5cm}

\begin{table*}
\caption{Elemental abundances $^{\dag}$} 
\label{tab:ab}
\begin{tabular}{lcccc}

\hline\hline
   Element & SMC N77 & mean SMC$^{\ddag}$  & LMC N191 & mean LMC$^{\ddag}$ \\
\hline
He/H                & 0.079 & 0.081 & 0.070 & 0.089\\
O/H (\x\,$10^{4}$)  & 1.04  &  1.07 & 2.0 & 2.24\\
N/H (\x\,$10^{6}$)  &  3.85 &  4.27 & 8.58 & 13.8\\
Ne/H (\x\,$10^{5}$) &  1.34 &  1.86 & 1.09 & 4.07\\
\hline 
\end{tabular}

$\dag$ See Sect. 3.5 for uncertainties \\
$\ddag$ \citet[][]{Russell92}

\end{table*}

\section{Discussion}

The two \hii\ regions studied in this paper, LMC N191A and SMC N77A, belong to 
the class of compact \hii\ regions, which are regions of newly
formed massive stars in the Magellanic Clouds.  
Their sizes (\ab\, 5 to 10\frac\ ) are much smaller than those of typical \hii\ 
regions in the Magellanic Clouds (several arc minutes). They are also associated with a 
much smaller number of exciting stars. With an \oiii\,(\lam \lam\,4959\,+\,5007)\,/\,\hb\ ratio 
of \ab\,2 and an \hb\ luminosity of 8.3\,\x\,10$^{36}$ erg s$^{-1}$, 
LMC N191A is a low-excitation blobs (LEBs), as defined by 
\citet[][]{Meynadier07}. 
At the same \hb\ luminosity, LEBs have lower excitation than HEBs and are powered by less 
massive exciting stars. 
In contrast, SMC N77A conforms more to the 
defining criteria of  high-excitation blobs (HEBs), because it has an 
\oiii /\hb\ ratio of \ab\,4 and an \hb\ luminosity of 1.1\,\x\,10$^{36}$ erg s$^{-1}$. 
Nevertheless, with a diameter of \ab\,20\frac , N77A is more extended than a typical HEB. 
Compared with N191A, N77A is more than a factor of two larger,  
about 40\% more massive, but less dense and less extincted (Table 1). \\

N191 is situated outside of the main body of 
the LMC, at a large distance (3 kpc) from the major  
star-forming region 30 Dor. Among the southernmost \hii\ region of LMC only 
N214 \citep[][and references therein]{Meynadier05}  
and N206 \citep[][and references therein]{Romita10}  
have been investigated in detail. In comparison with these two complexes, 
N191 is a smaller region with a fainter emission nebula. 
It is also linked to an OB association, LH 23  \citep[][]{Lucke70}, and a giant 
molecular cloud, 54\,\x\,14 pc in size with a CO mass of 2\,\x 10$^{5}$ \sm\ \citep[][]{Fukui08}.   
N77 is the northernmost \hii\ region of the SMC so far studied in detail. 
It lies some 440 pc north of N66, the main star-forming region in the SMC 
 \citep[][and references therein]{MHM10a}. N77 is associated with a small 
molecular cloud   \citep[][]{Mizuno01} and a small OB association, B-OB 24 
\citep[][]{Battinelli91}. \\

An accurate characterization of the exciting source of each of these two compact 
\hii\ regions requires a more detailed investigation. There is indeed a discrepancy between 
the spectral type indicated by the spectra and that derived from \hb\ flux estimates. 
The spectrum obtained toward star \#1 in LMC N191A belongs to an O8.5 V type (Sect. 3.4). 
In contrast, the \hb\ flux indicates an earlier O5 V type (Sect. 3.3). 
The same is true for SMC N77A. The spectral classification indicates an early B type star 
(Sect. 3.4), whereas the \hb\ luminosity suggests an O8 V type at least (Sect. 3.3). 
This discrepancy can be accounted for 
by the presence of hotter stars embedded in the \hii\ regions. 
This assumption is in line with the indication of the ($V, B-V$) color-magnitude diagram 
(Fig.\,\ref{fig:cmd_n191}) that is  a 40 \sm\ star of 
3 Myr old for the main exciting source of N191A. For N77A, we detected 
the nebular \oiii\ line \lam\,4363, which needs a much earlier exciting star than 
a B type. We note, however, that this line is not reported in the paper by 
\citet[][]{Russell90}. Still, high-resolution observations in near-IR are 
necessary to check the possibility of embedded sources. Another explanation 
is that the spectral type is underestimated because of 
contamination from nebular emission 
lines. However, justifying a three-subtype uncertainty seems difficult. \\

The difference between the excitation degrees of LMC N191A and SMC N77A, as mentioned 
above, can be commented upon from another viewpoint: 
the \oiii /\hb\ ratio is higher in N77A (\ab\,4) compared to that 
in N191 (\ab\,2), even if the latter is denser. According to models for homogeneous \hii\ regions 
\citep[e.g.,][]{Stasinska90} the  \oiii /\hb\ ratio is proportional to the electron 
density for a given exciting source. This means that we expect a lower ratio 
for N77A if the exciting sources have comparable effective temperatures. The higher 
 \oiii /\hb\ ratio in N77A would suggest a hotter star than in N191A. 
Alternatively, the weak ratio of N191A may be due to the density structure 
of N191A. The aforementioned models predict that a density rise in the 
outer zones of an \hii\ region results in a decrease of the global \oiii /\hb\ ratio. 
Otherwise the higher \oiii /\hb\ 
ratio of N77A may reflect the difference of metallicity between the LMC and SMC. 
In low-metallicity environments, the inefficiency of
cooling raises the electron temperature so that forbidden oxygen lines
become stronger despite the lower abundance. These assumptions may explain 
the apparent discrepancy  
 between the higher \oiii /\hb\ ratio of N77A and the cooler ionizing star inferred from 
our study.\\

Three young stellar object (YSO) candidates detected by 
\citet{Gruendl09} lie toward N191A. Similarly, there are three such candidates 
detected in the N77 field \citep{Bolatto07}. They are indicated in 
Figs.\,\ref{fig:n191} and \ref{fig:n77} (lower panels) and 
Figs.\,\ref{fig:n191_spitzer} and  \ref{fig:n77_spitzer}, respectively. 
Their coordinates are listed in Table\,\ref{tab:phot_yso}. 
The N191 candidates lack an optical counterpart except for object \#13. The closest 
candidate to N191A, named 050435.85-705430.1 (or N191-YSO1 in the present paper), 
lies about 15\frac\ (3.6 pc) northeast of the exciting star \#1. The {\it JHK} and Spitzer 
IRAC and MIPS colors of these objects are listed in Table 3.  
Two of the N77 candidates (YSO2 and YSO3) seem very close to two 
relatively bright field stars, which raises the question of their association. 
However, astrometrically speaking, these stars and the YSO candidates are not 
associated. It seems that these YSO candidates have very faint optical 
counterparts below our detection limits.\\

A comprehensive search for YSOs in the LMC has also been carried out by the SAGE team 
and was reported by \citet{Whitney08}.
They have found only one YSO in the vicinity of N191A that corresponds to the closest 
candidate of \citet{Gruendl09}. 
This difference is due to the different selection criteria, based on IRAC colors, 
used in different works.  
\citet{Gruendl09} argue that there are no simple criteria in color-magnitude 
space that can unambiguously separate the YSOs 
from AGB/post-AGB stars, planetary nebulae, and background galaxies.
Moreover, the point source definition differs in the two approaches. 
\citet{Gruendl09} include slightly extended sources that are likely 
YSOs superimposed on a bright background. In contrast, because the SAGE definition 
is more constraining, it excludes such cases.   \\ 

To be more specific, here we applied the selection diagrams used by 
\cite{Simon07} to look into the nature of the YSO candidates. These authors 
used color-color diagrams [3.6]-[4.5] versus [5.8]-[8.0] and 
[3.6]-[4.5] versus [4.5]-[8.0] to characterize 
the candidate YSOs they have detected toward the SMC \hii\ region NGC 346 (N66). 
We note that on the [3.6]-[4.5] versus [5.8]-[8.0] plot two of the YSO 
candidates have colors near to those of YSOs. These are N191 YSO3 and N77 YSO1. The 
other candidates show colors of ``probable YSOs'' or ``poor fits/PAHs''. 
Note that the [3.6]-[4.5] versus [5.8]-[8.0] plot does not clearly separate YSOs 
from other types of sources, 
particularly stars with modest IR excesses and sources with PAH contamination 
\citep{Simon07}. We also used  
the [3.6]-[4.5] versus [4.5]-[8.0] plot, which takes advantage of longer color baselines and the 
abrupt change in YSO spectra between the 4.5 and 5.8 $\mu$m bands, to distinguish YSOs 
from stars, galaxies, and PAH. However, on this plot all YSO candidates are offset 
with respect to the expected YSO positions, since they are redder, i.e. 
with higher [4.5]-[8.0] color values, compared to YSOs.   \\

A spectral energy distribution (SED) analysis provides a more efficient method 
for investigating the nature of YSOs. We used the Spitzer photometry to construct the 
mid-IR SEDs of our YSO candidates. We fitted these SEDs with the library of YSO models 
by \cite{Robitaille06} using the online SED fitting tool provided by these authors 
\citep{Robitaille07}\footnote{ and available at 
http://caravan.astro.wisc.edu/protostars}.  
The SED plots of the brightest YSO candidates toward N191 and N77 are displayed in 
Fig.\,\ref{fig:yso}, N191-YSO1 is best fitted by models of 16 to 21 \sm . 
The best fit suggests a 20 \sm\ protostar with a total luminosity 
of 2.9\,\x\,10$^{4}$ \slum . 
With regard to  N77-YSO1, the best models belong to  masses ranging from 
10 to 17 \sm . The best fit is for a 10 \sm\ protostar with a  total luminosity of 
1.0\,\x\,10$^{4}$ \slum . The majority of the best-fit YSO models include both 
circumstellar envelopes and disks. 
Moreover, in these models the accretion rate from the envelope onto the 
YSO indicates the evolutionary stage of the protostar \citep{Robitaille06}. 
Based on the high accretion rates, 1\,\x\,10$^{-4}$ and 5\,\x\,10$^{-5}$ \sm\ yr$^{-1}$ for 
N191-YSO1 and N77-YSO1, respectively, 
both objects can be classified as Stage I sources \citep{Robitaille06}.  
This classification is equivalent to the traditional Class I source. 
It should, however, be cautioned that these models are based on low-mass star formation 
scenarios, whereas we do not know how massive stars actually form. Therefore,   
massive YSOs may contrast in their properties with commonly studied low-mass YSOs, in particular 
in low-metallicity environments such as the Magellanic Clouds.    
However, the use of these models must be considered as a first approach and preliminary 
screening of the problem. \\

The two brightest YSO candidates are also the most closely adjacent objects to their respective 
\hii\ regions. More specifically, N191 YSO1 lies 15\frac\ northwest of star \#1, while 
N77 YSO1 is seen toward the central dust lane of the compact \hii\ region, very 
close to the exciting star. Since N191 and N77 are young active \hii\ regions, these 
two YSO candidates may effectively be associated with them.   
The presence of these YSO candidates confirms  that star formation activity is still 
ongoing in N191A and N77A. More specifically, massive protostars of \ab\, 10 and 
20 \sm\ are in the process of formation. 
The YSOs may have been triggered by the ionization front progression in the associated  
molecular clouds. However, high-resolution submillimeter observations, such as 
those of ALMA, are required to check these first results. \\

\section{Concluding remarks}

This paper presented the first detailed study of LMC N191A and SMC N77A using 
imaging and spectroscopy in the optical obtained at the ESO NTT as well as Spitzer 
and 2MASS data archives. The two objects are among the outermost star-forming regions of the 
Magellanic Clouds. We derived several physical characteristics of these regions and their 
powering sources. The compact \hii\ region N191A, \ab\,10\frac\ (2.4 pc) in diameter,  
belongs to a small class of ``low-excitation blobs'' 
in the Magellanic Clouds. In contrast, SMC N77A, \ab\,20\frac\ (5.8 pc) in size,  
belongs to the ``high-excitation blob''  family. 
The class of compact \hii\ regions in the Magellanic Clouds is not very populated. Therefore new 
members provide additional data for improving our knowledge of their 
characteristics and their formation processes.  
Higher resolution observations are necessary to deepen the study of these objects.

\end{document}